# Enlarging the magnetocaloric operating window of the Dy$_2$NiMnO$_6$ double perovskite

M. Balli[1,2] *, S. Mansouri[2], P. Fournier[2,3], S. Jandl[2], K. D. Truong[2], S. Khadechi-Haj Khlifa[4], P. de Rango[4], D. Fruchart[4], A. Kedous-Lebouc[5]

[1] LERMA, ECINE, International University of Rabat, Parc Technopolis, Rocade de Rabat-Salé, 11100, Morocco.

[2] Institut Quantique & Département de physique, Université de Sherbrooke, J1K 2R1, QC, Canada.

[3] Canadian Institute for Advanced Research, Toronto, Ontario M5G 1Z8, Canada.

[4] Institut Néel & Université Grenoble Alpes, B.P. 166, Cedex 9, Grenoble, France

[5] G2Elab, Grenoble Institute of Technology, 21 avenue des martyrs, 38031 Grenoble CEDEX1, France

*mohamed.balli@uir.ac.ma

**Abstract**

In this paper, we mainly focus on the magnetic and magnetocaloric features of La$_{2-x}$Dy$_x$NiMnO$_6$ double perovskites. Their magnetocaloric properties are investigated in terms of both entropy and adiabatic temperature changes. In contrast to early works, it was found that the Dy$_2$NiMnO$_6$ compound unveils dominant antiferromagnetic interactions under very low magnetic fields. The ordering of its Dy$^{3+}$ magnetic moments is associated with a giant magnetocaloric effect at very low temperatures, while the established ferromagnetic Ni-O-Mn super-exchange interactions close to 100 K give rise to a moderate magnetocaloric level, only. On the other hand, the doping of Dy$_2$NiMnO$_6$ with high amounts of large-size rare earth elements such as La would enable us to cover an unusually wide magnetocaloric temperature range going from the liquid helium temperature up to room-temperature. More interestingly, the presence of both ordered and disordered ferromagnetic phases in La$_{1.5}$Dy$_{0.5}$NiMnO$_6$ maintains constant the isothermal entropy changes over a temperature span of about 200 K, being a favorable situation from a practical point of view.

## I. Introduction

The search for new advanced magnetocaloric materials is highly important for the development of magnetic refrigeration technology. The latter technique was proven to be more efficient and ecofriendly when compared to conventional cooling [1-3]. With the purpose of implementing magnetic refrigerants other than gadolinium and its related alloys, research activities have been markedly stimulated during the last two decades leading to the discovery of interesting compounds from both performance and economical points of view [1]. This mainly includes some intermetallics containing cheaper elements with NaZn$_{13}$ and Fe$_2$P structures that usually exhibit a giant magnetocaloric effect (MCE) close to room-temperature[1]. However, although these materials were successfully tested in numerous magnetic refrigeration prototypes, their poor chemical resistance against oxidation and corrosion phenomena as well as their mechanical brittleness would restrict their utilization as refrigerants in particular devices [1, 4]. This is because the considered magnetocaloric candidate is expected to experience billions of thermodynamic active magnetic refrigeration cycles (AMR) [1-3] during its working life, involving direct contacts with heat transfer fluids such as water. On the other hand, up to now the reported research works in relation with the reference $NaZn_{13}$ and $Fe_2P$-based compounds fail to improve their performance particularly in terms of the adiabatic temperature change that remains comparable or even lower to that of the reference gadolinium metal [1], opening the way for further investigations in the field.

In functional magnetocaloric devices, the manganites-based materials [5, 6] show some key assets when compared to intermetallics such as high chemical resistance against oxidation and corrosion phenomenon, strong mechanical stability and, high electrical resistance (absence of eddy currents). This would compensate for their relatively lower magnetocaloric features. Additionally, the magnetic frustration phenomena shown by several manganites such as $RMnO_3$ and $RMn_2O_5$ (R = Rare Earth) [6, 7-8] enables us to generate MCEs by simply rotating them in constant magnetic fields instead the conventional magnetization-demagnetization process. More recently, materials such as $R_2MM'O_6$ (R = lanthanide, M and M' = transition metals) double-perovskites begin to attract an increasing interest not only for their potential applications in spintronic devices [9] but also because of their significant magnetocaloric properties [10-16]. It was known that this family of manganites unveils several magnetic, electric and structural ordering parameters resulting in a wide variety of intriguing and fascinating physical properties [9-23]. Taking advantages of this multiferroicity, different tasks such as refrigeration and data storage could be achieved by simply using a single $R_2MM'O_6$ material which is highly appreciated from both environmental and economical points of view. Also, the $R_2MM'O_6$ magnetic properties are usually driven by super-exchange interactions involving M-O-M' bonds which are very sensitive to any structural and\or electronic changes [14-25]. Consequently, their magnetocaloric properties can be easily tailored under pressure or by doping the rare earth and manganese sites.

In the work by Rogado et al [9], it was reported that the dielectric constant of $La_2NiMnO_6$ double perovskite strongly depends on the magnetic field magnitude. For example, the phase transition exhibited by the dielectric constant around 220 K in the absence of magnetic field, can be easily shifted up to 280 K when the compound is subjected to a very small magnetic field of only 0.1 T. This underlines a strong interplay between the magnetic and dielectric properties near room temperature paving the way to the design of new spintronic devices. More recently, Balli et al [15, 16] have studied the potential of ferromagnetic $La_2(Ni, Co)MnO_6$ compounds in magnetic refrigeration. More particularly, it was found that the ordered phase of $La_2NiMnO_6$ single crystals presents a large refrigerant capacity around room temperature which compares well with that shown by some giant MCE materials [15]. In addition, the transition temperature of the $La_2NiMnO_6$ double perovskite can be easily shifted down to temperatures around 200 K by only changing synthesis conditions resulting in a wide operating temperature range [14, 15]. By using the thin films approach, Matte et al [14] have demonstrated that an unusual temperature range comprised between 100 and 300 K can be covered by using only a single $La_2NiMnO_6$ compound without need to magnetocaloric multilayers [1]. This was made possible thanks to the control of the proportions of ordered and disordered ferromagnetic phases through the growth pressure and the annealing temperature [14].

Usually, the well-ordered $La_2NiMnO_6$ double perovskites have a Curie temperature closer to 280 K [15, 24-25]. However, according to a recently reported work by Ravi and Senthikumar [26], the Curie temperature shown by $La_2NiMnO_6$ nanoparticles reaches a record value of about 350 K. Such unusual $T_C$ was attributed by the authors to oxygen-induced defects [26]. More recently, the magnetocaloric properties of $R_2(Ni, Co)MnO_6$ double perovskites with magnetic rare earth elements ( R = Tb, Gd, Ho,

Dy, Pr, Er and Nd) were also investigated [10-13]. For the compounds $R_2NiMnO_6$ and $R_2CoMnO_6$ with R = Dy, Ho, Er, the Curie temperature arising from 3d transition metals ranges between 70 and 100 K, while the rare earth ordering temperature is below 10 K [10, 11]. All these materials exhibit a large MCE at very low temperatures (around 5 K) that can attain a maximum entropy change of about 11 J/kg K under a magnetic field change of 7 T [10, 11]. In similar conditions of temperature range and magnetic field, the $Gd_2NiMnO_6$ and $Gd_2CoMnO_6$ double perovskites intriguingly reveal a maximum entropy change overpassing twice those shown by $R_2NiMnO_6$ and $R_2CoMnO_6$ (R = Dy, Ho, Er) compounds [13]. Regarding the $R_2NiMnO_6$ compounds with R = Tb, Nd and Pr, larger Curie temperatures of 110, 191 and 213 K were reported, respectively. Their associated maximum entropy changes were found to be comprised between 2 and 4 J/kg K [12].

In this paper, we mainly investigate the magnetic and magnetocaloric properties of $La_{2-x}Dy_xNiMnO_6$ double-perovskites. By a such study we aim to take advantage of the numerous magnetic phase transitions involving both 4f and 3d elements. A situation that enables us to significantly enhance the working magnetocaloric temperature range of this family of materials. We are also aiming to figure out how the substitution of Dy by large size lanthanides such as La influences the magnetic and magnetocaloric features of $Dy_2NiMnO_6$.

## II. Experimental

Polycrystalline samples of $La_{2-x}Dy_xNiMnO_6$ compounds were prepared by the solid-state reaction technique using stoichiometric amounts of high purity starting materials of $La_2O_3$, $Dy_2O_3$, NiO and $Mn_2O_2$ following the reported procedure by Truong et al in Ref.27. First, the $La_2O_3$ and $Dy_2O_3$ were pre-calcined in air at 900 °C. Stoichiometric amounts of these oxides were then mixed, ground and poured into an alumina crucible. The mix was subjected to several heating cycles between 1000 and 1300 °C in air with various intermediate grinding steps to obtain a homogeneous mixture. In the final step, the resultant $La_{2-x}Dy_xNiMnO_6$ powder was pressed into pellets and annealed at 1450 °C in air overnight. The quality and the crystalline structure of studied samples were checked by X-ray diffraction (XRD) powder measurements performed at room temperature using a Bruker-AXS D8-Discover diffractometer with the $CuK\alpha1$ radiation that allows different scan modes. The analysis of XRD patterns for $La_{2-x}Dy_xNiMnO_6$ (x = 0.5 and 2) unveils that the studied samples crystalize rather in an orthorhombic structure (weakly monoclinic) with the *Pbnm* space group. Their unit cell parameters were found to be a = 5.505, b = 7.747 and c = 5.452 Å for $Dy_2NiMnO_6$, and a = 5.51, b = 7.77 and c = 5.46 Å for $La_{1.5}Dy_{0.5}NiMnO_6$. Magnetic measurements under magnetic fields changing from 0 up to 7 T in the temperature range 2-300 K were performed using a superconducting quantum interference device (SQUID) from Quantum Design, model MPMS XL. The homemade BS1 and BS2 magnetometers developed at Néel Institute, Grenoble, were also employed to collect magnetization data. Specific heat measurements were carried out as a function of temperature (from 2 to 350 K) at zero magnetic field with a Quantun Design Physical Properties Measurements system (PPMS) using a relaxation method.

## III. Results and discussions

The temperature dependence of the magnetization measured on the $Dy_2NiMnO_6$ compound under the low magnetic field of 0.1 T in zero-field cooled (ZFC) and field-cooled (FC) modes is plotted in Figure 1-a. With increasing temperature, several anomalies can be clearly observed at 5, 46 and 98 K. The $T_C$ = 98 K feature can be ascribed to the occurrence of a magnetic phase transition from paramagnetic to ferromagnetic state due to the $Ni^{2+}$ ($t^6_{2g} e^2_g$)-O (2p)-$Mn^{4+}$ ($t^3_{2g} e^0_g$) superexchange interactions. The latter involve 3d metals exchanges between empty ($Mn^{4+}$) and half filled ($Ni^{2+}$) eg orbitals that are mediated by oxygen atoms, following the Goodenough-Kanamori-Anderson rules [28-30]. Their strength and nature (negative, positive) are determined by the length and the angle of the M-O-M'($Ni^{2+}$-O-$Mn^{4+}$) bond. The anomaly at $T_R$ = 5 K is more probably attributed to the ordering of $Dy^{3+}$ magnetic moments with $Dy^{3+}$-O-$Dy^{3+}$ exchange interactions. Finally, the magnetic transition taking place at $T_{R-d}$ = 46 K may originate from a partial polarization of $Dy^{3+}$ moments thanks to 3d-4f exchange interactions. The present $T_R$ and $T_C$ values are well consistent with those previously reported ($T_R$ = 6 K and $T_C$ = 101 K) in Ref.10. In contrast, our obtained $T_{R-d}$ is much larger than that reported by Jia et al (19.5 K) [10]. The reason behind that marked deviation remains unclear which requires further investigations.

The 0.1 T-ZFC reciprocal magnetic susceptibility (1/χ) as a function of temperature is reported in Figure 1-b. As shown, the linear feature of 1/χ at high temperatures above $T_C$ reveals that the 1/χ (T) curve follows the Curie-Weiss law χ = C/(T-$T_\Theta$), where C is the Curie-Weiss constant. By fitting the experimental 1/χ with ideal Curie-Weiss curve, the effective magnetic moment and the paramagnetic Curie-Weiss temperature ($T_\Theta$) were found to be 16.8 $\mu_B$ and -24 K, respectively. The deduced effective magnetic moment is in perfect agreement with the theoretically expected value given by $\mu_{eff} = \sqrt{(\mu_{eff}(2*Dy^{3+}))^2 + (\mu_{eff}(Ni^{2+}))^2 + (\mu_{eff}(Mn^{4+}))^2} = 16.525 \mu_B$. The effective magnetic moments of $Dy^{3+}$ (10,65 $\mu_B$), $Ni^{2+}$(3,873 $\mu_B$) and $Mn^{4+}$(5,59 $\mu_B$) were taken from Ref.34 which are calculated assuming a spin-orbit coupling. In this paper, we assume that the magnetic moments are mainly determined by the oxidation state of Mn and Dy ions. This seems to be confirmed by the experimental and calculated effective magnetic moments. However, the superexchange interactions between B-site cations and oxygen ions as well as the LS-coupling in lanthanides should be also considered. On the other hand, the negative value of $T_\Theta$ clearly indicates dominant antiferromagnetic interactions in the $Dy_2NiMnO_6$ double perovskite. This markedly contrasts with Jia et al study [10] in which a positive value of $T_\Theta$ (38 K) was reported, suggesting ferromagnetic interactions. In fact, as demonstrated for the $Nd_2NiMnO_6$ compound [31], the strong ferromagnetic (ferrimagnetic) interaction involving the Ni-O-Mn bond would polarize the $Dy^{3+}$ magnetic sublattice via a negative 3d-4f exchange interactions. The new established interactions lead to a canted antiferromagnetic arrangement of $Dy^{3+}$ magnetic moments with respect to the $Ni^{2+}$ and $Mn^{4+}$ ferromagnetic-sublattices. A similar behaviour was recently observed in other $R_2NiMnO_6$ double perovskites [32, 33]. This may explain the relatively low magnetic moment of $Dy_2NiMnO_6$ even under high magnetic fields as found at low temperatures. For example, at 6 K, the obtained magnetization in the magnetic field of 7 T is 13.2 $\mu_B$ being only about 50 % of the expected saturation magnetization which is approximately 25.5 $\mu_B$, considering contributions from both $Ni^{2+}$-O-$Mn^{4+}$ (~ 5.5 $\mu_B$) [15] and Dy (10 $\mu_B$) [34] magnetic moments. However, grain boundary effects would also contribute for this loss in terms of the saturation magnetization [5].

The characterization of magnetocaloric materials is usually carried out based on the entropy (ΔS) and adiabatic ($\Delta T_{ad}$) temperature changes that can be simultaneously evaluated from specific heat data. However, as the calorimetric measurements are highly challenging, the MCE is frequently reported in terms of ΔS that can be easily determined from magnetic isotherms with the help of the well-known Maxwell relation [1] given by $\Delta S (T, 0-H) = \mu_0 \int_0^H \left(\frac{\partial M}{\partial T}\right)_{H'} dH'$. In the absence of large hysteretic effects ~~and~~ leading to phase-separated states [35], this relation enables us a fast and efficient assessment of magnetocaloric materials. In addition, the above equation is also appropriate for the determination of ΔS in magnetic materials presenting second order magnetic phase transitions (SOMT)[1]. The here studied compounds exhibit a negligible hysteresis effect as can be seen in magnetization loops reported for $Dy_2NiMnO_6$ in Figure 2, at two representative temperatures (for example). On the other hand, the positive slopes shown by Arrott plots (H/M vs $M^2$) (not shown here)[36] indicate that both $La_{2-x}Dy_xNiMnO_6$ (x = 0.5 and 2) compounds unveil a SOMT. At first, the magnetization method was utilized to evaluate the $Dy_2NiMnO_6$ entropy change which is plotted in Figure 3-a as a function of temperature for several magnetic fields. As shown, only the peaks associated with the $Ni^{2+}$-O-$Mn^{4+}$ super-exchange interactions and the ordering of $Dy^{3+}$ magnetic moments can be clearly identified close to 100 and 5 K, respectively. The ΔS at $T_R$ is at least 5 times larger than that shown by $Dy_2NiMnO_6$ near its Curie point. In the magnetic field change of 5 T, a maximum entropy change of 11 J/kg K can be obtained at 5 K and only 1.85 J/kg K at 100 K. This is partly due to the large change in the $Dy^{3+}$ ions magnetization as compared to $Ni^{2+}$ and $Mn^{4+}$ ions.

It is worth noting that our reported ΔS values significantly differ from those reported by Jia et al for a similar double perovskite [10]. For a magnetic field changing from 0 to 5 T, Jia et al have obtained maximum ΔS values of 9 J/kg K at 6.5 K and 3 J/kg K at 100 K. The observed deviation could be attributed to different reasons such as synthesis conditions, grains size of prepared samples and the presence of additional minor phases. It is also worthy to note that although the $Gd_2NiMnO_6$ compound contain a rare earth element (Gd) with lower magnetic moment (7 $\mu_B$) [34], it was reported that its maximum entropy change (36 J/kg K under 7 T) in the temperature range around 6 K, intriguingly exceeds that of $Dy_2NiMnO_6$ at least by a factor of 3 [13]. However, the physics behind this marked enhancement in the MCE remains not understood.

The isothermal entropy change indicates on the cooling energy that can be provided by magnetocaloric materials when subjected to an external magnetic field. However, the adiabatic temperature change is also of great interest since it determines the strength of heat exchanges in magnetic refrigerators. For the present $Dy_2NiMnO_6$ compound, the $\Delta T_{ad}$ was determined by combining magnetization and zero-field specific heat data reported in Figure 4. Because the $R_2M'MnO_6$ specific heat slightly depends on the applied magnetic field [16], the adiabatic temperature change can be approached by $\Delta T_{ad} = -\frac{T}{C_P}\Delta S$ [1]. The latter is shown in Figure 3-b as a function of temperature under a magnetic field change of 5 T. As shown, $\Delta T_{ad}$ exhibits two maximums of about 8 K around 5 K and only 1 K around 100 K, corresponding to the ordering of $Dy^{3+}$ and Ni-O-Mn magnetic couplings, respectively. The low value of $\Delta T_{ad}$ around 100 K is not only attributed to the moderate entropy change in this temperature range (Fig.3-a), but also to the large value of the specific heat (Fig. 4) which is dominated by phonons contributions at high temperatures.

As reported above, the $Dy_2NiMnO_6$ compound unveils significant MCE levels at temperatures below 100 K. However, for potential room-temperature tasks, it is necessary to shift these thermal effects toward 300 K. Fortunately, it was shown that in $R_2NiMnO_6$-double perovskites, the Curie temperature increases almost linearly with the atomic radius of the rare earth element ($r_{R3+}$) to attain the room temperature region for the largest $R^{3+}$ ions [24, 25]. This is associated to modifications in the superexchange angle of the Ni-O-Mn bond. As the latter increases to a larger value with increasing the lanthanide size, the resulting ferromagnetic interactions are enhanced and accordingly the Curie temperature, following the Goodenough-Kanamori-Anderson rules [28-30, 24, 25]. Keeping in mind this linear relationship between $T_C$ and $r_{R3+}$, the compound $Dy_2NiMnO_6$ was doped with lanthanum following the $La_{1.5}Dy_{0.5}NiMnO_6$ formula. This enables us to partly enhance the Ni-O-Mn ferromagnetic interactions while taking advantage of the remaining magnetic interactions arising from the $Dy^{3+}$ sublattice. Consequently, the number of magnetic phase transitions would be increased leading to a large working magnetocaloric temperature range. In Figure 5-a, we report the FC and ZFC-temperature dependence of magnetization for the $La_{1.5}Dy_{0.5}NiMnO_6$ compound under a low magnetic field of 0.1 T. As shown, the modified $Dy_2NiMnO_6$ unveils pronounced magnetic transitions close to 255 and 130 K. The observed high $T_C$ at 255 K indicates the presence of the partial atomically ordered $Ni^{2+}/Mn^{4+}$ resulting in the ferromagnetic $Ni^{2+}$ ($t^6_{2g}$ $e^2_g$)-O (2p)-$Mn^{4+}$ ($t^3_{2g}$ $e^0_g$) super-exchange interactions. The low transition temperature phase around 130 K is more probably attributed to the trivalent oxidation state $Ni^{3+}/Mn^{3+}$, suggesting an incomplete ordering of Ni and Mn atoms [14]. As can be observed, the magnetic phase transition involving the $Dy^{3+}$ ions, is not clearly visible from the thermomagnetic curve reported in Figure 5-a. This is mainly due to the weak polarization of $Dy^{3+}$ magnetic moments under low magnetic fields (0.1 T). However, the ordering of $Dy^{3+}$ magnetic sublattice can be clearly observed under sufficiently high magnetic fields (not shown here). The Curie-Weiss paramagnetic temperature is found to be 217 K indicating dominant and strong ferromagnetic interactions in the $La_{1.5}Dy_{0.5}NiMnO_6$ compound, which markedly contrast to the present studied $Dy_2NiMnO_6$. Assuming an equal contribution of $Ni^{2+}/Mn^{4+}$ and $Ni^{3+}/Mn^{3+}$ oxidation states, the $La_{1.5}Dy_{0.5}NiMnO_6$ theoretical effective magnetic moment was obtained to be 9.97 $\mu_B$, being in very good agreement with the experimental value (9.87 $\mu_B$) deduced from the reciprocal magnetic susceptibility (Fig. 5-b). Once again, the effective magnetic moments of $Ni^{3+}$ (3,872 $\mu_B$) and $Mn^{3+}$ (4,899 $\mu_B$) were taken from Ref. 34 assuming a spin orbit interplay. The reduction of $La_{1.5}Dy_{0.5}NiMnO_6$ effective magnetic moment mainly arises from the dilution of the $Dy^{3+}$ magnetic sublattice.

It is worth noting that in addition to structural parameters, the superexchange couplings can be also affected by the crystal field. For the $Ni^{2+}/Mn^{4+}$ oxidation state, the Hund's interplay between $t_{2g}$ and eg states of Mn results in a parallel alignment of electronic spins. In addition, the $t_{2g}$ states of Ni are fully occupied. As a result of the Hund's coupling, the remaining two electrons occupy the eg states with a parallel configuration. The latter contribute to the superexchange coupling via O-p state leading to a preferred spin-up Ni eg electrons configuration. Such situation explains the high ordering temperature ferromagnetic interactions for $Ni^{2+}/Mn^{4+}$ oxidation state [37]. However, for the $Ni^{3+}/Mn^{3+}$ oxidation state, the established superexchange coupling would have result in antiferromagnetic ordering. This is not the case because of the strong crystal field splitting on Mn ions [37]. For more details about the electronic structure of $La_2NiMnO_6$-based double perovskites, we refer the interested reader to our recently published work in Gauvin Ndiaye et al [37].

The presence of the disordered magnetic phase in addition to the ordered one in the double perovskite $La_{1.5}Dy_{0.5}NiMnO_6$ is not a drawback to be avoided since it allows us to extend the magnetocaloric operating window over a large temperature range [14]. For example, in Figure 6-a we report the isothermal entropy change as a function of temperature for $La_{1.5}Dy_{0.5}NiMnO_6$ under the magnetic field change of 7 T. As shown, ΔS remains practically constant over a wide temperature span going from room temperature region down to 50 K. This sort of behaviours is highly appreciated from a practical point of view, particularly in cases where the cooling process is achieved by using the Ericson cycle [1]. In such a cycle, the difference between the full entropy curves at zero and non-zero magnetic fields must be maintained unchanged between the cold and hot ends, being a necessary condition for reaching the theoretical limit of Carnot efficiency [1, 38]. On the other hand, aiming to generate a large temperature span between hot and cold sources, the most majority of magnetic cooling devices employs the active magnetic regeneration refrigeration (AMRR) method which is constituted of a cascade of Brayton cycles [38, 39]. Usually, to meet the AMRR and Ericsson cycles requirements, several materials with distributed Curie temperatures in the considered temperature range are combined in an optimum way to form a multilayered-refrigerant [1, 40, 41]. This solution would involve additional costs and engineering difficulties, while the same task can be achieved by simply employing a single double perovskite material as demonstrated in Figure 6-a. By using the thin films approach, we have more recently demonstrated that a usually "table-top" like isothermal entropy change can be generated between 300 and 200 K through the control of ordered and disordered phases amounts in the $La_2NiMnO_6$ compound, via growth conditions [14]. In this paper, we report evidences showing that it is possible to obtain a similar behaviour using bulk forms. In addition, as reported in Figure 6-a, the doping of $La_2NiMnO_6$ by a relatively small amount of magnetic rare earths such as Dy allows to markedly extend the operating temperature range down to very low temperatures because of the $Dy^{3+}$ magnetic moments ordering close to 10 K. However, the exhibited moderate entropy change over this large temperature range particularly at high temperatures remains a serious obstacle behind the practical implementation of this kind of materials, opening the way for further investigations.

In Figure 6-b, the $La_{1.5}Dy_{0.5}NiMnO_6$ adiabatic temperature change is plotted as a function of temperature under a magnetic field variation of 7 T. As shown, the $\Delta T_{ad}$ remains approximately constant over the temperature range between 50 and 300 K revealing an unusual behaviour when compared to that of standard magnetocaloric materials. This behaviour is due to the fact that the $La_{1.5}Dy_{0.5}NiMnO_6$ specific heat almost linearly increases with temperature (Fig. 4) leading to a nearly constant $T/C_p$ while the entropy change remains approximately unchanged for temperatures above 50 K. In the magnetic field change of 7 T, the $La_{1.5}Dy_{0.5}NiMnO_6$ compound exhibits a maximum $\Delta T_{ad}$ of about 6.75 K at 6 K, while around 250 K only a temperature change of 0.5 K can be achieved under a similar magnetic field variation.

Looking at Figures 3 and 6, it clearly seems that the $La_{1.5}Dy_{0.5}NiMnO_6$ oxide cannot be directly implemented in functional devices because of its moderate MCEs. The latter are partly attributed to the grain boundary effects that are common features of polycrystalline manganites [5, 16]. However, the $La_{1.5}Dy_{0.5}NiMnO_6$ magnetocaloric properties can be markedly enhanced by increasing the grains size of its polycrystalline forms using optimum annealing conditions [42-44]. This increase in the grains volume/surface ratio would enable us to reduce the contribution from the magnetic disordered surface boundary [42, 43]. Additionally, when considering the case in which the fully disordered magnetic moments of Dy, Ni and Mn atoms are completely oriented parallel to the applied magnetic field, the resulting entropy change is expected to reach the theoretical limit given by $\Delta S_{Limit} = R*Ln(2J+1) = 40$ J/kg K, where R is the universal gas constant and J is the effective angular momentum quantum number. For $La_{1.5}Dy_{0.5}NiMnO_6$, the J value (5) was deduced from the magnetization saturation given by $m_0 = g*J*\mu_B \approx 10\ \mu_B$, with the Landé factor g assumed to be 2. However, the $\Delta S$ exhibited by $La_{1.5}Dy_{0.5}NiMnO_6$ under 7 T (for example) in the temperature range between 50 and 255 K is only 2.5 % of the theoretical limit. This opens the way to further fundamental and experimental investigations of $R_2NiMnO_6$ double perovskite with the aim to fully taking advantage of their magnetocaloric potential. Also, some of the $R_2NiMnO_6$ double perovskites exhibit additional degrees of freedom such as electrical polarization enabling potential supplemental caloric effects under electric fields. Finally, the high chemical stability of these family of multiferroics as well as their insulating character make them potential candidates for application in magnetocaloric devices.

**IV. Conclusions**

To sum up, we have particularly studied the magnetocaloric features of $(La_xDy_{1-x})_2NiMnO_6$ double perovskites. In contrast to early studies, it was observed that the $Dy_2NiMnO_6$ compound unveils dominant antiferromagnetic interactions under low magnetic fields. Its magnetocaloric effect exhibits two pronounced peaks around 5 and 100 K, corresponding to the ordering of $Dy^{3+}$ and $Ni^{2+}$-O-$Mn^{4+}$ magnetic couplings, respectively. Under a magnetic field changing from 0 to 5 T, the $Dy_2NiMnO_6$ adiabatic temperature change was found to reach a large maximum of about 10 K at very low temperature and only about 1 K around 100 K. The moderate value of $\Delta T_{ad}$ associated with the $Ni^{2+}$-O-$Mn^{4+}$ ferromagnetic transition is mainly due to the low value of the $Ni^{2+}$-O-$Mn^{4+}$ magnetization as well as the large value of the specific heat that is dominated by phonon contributions at high temperatures. On the other hand, the doping of $Dy_2NiMnO_6$ with large-size rare earth elements such as La enables us to markedly extend the operating magnetocaloric window from room-temperature range down to very low temperatures. More interestingly, the presence of ordered and disordered magnetic phases in $(La_xDy_{1-x})_2NiMnO_6$ allows to maintain constant the entropy change over a wide temperature range. For example, with the $La_{1.5}Dy_{0.5}NiMnO_6$ compound, the isothermal entropy change remains approximately unchanged over an unusual temperature range going from 50 up to 250 K, which is highly appreciated from a practical point of view. Even the adiabatic temperature change exhibits a similar behaviour, which was not observed before by any known magnetocaloric materials. However, the magnetocaloric properties of $(La_xDy_{1-x})_2NiMnO_6$ compounds need to be markedly improved before their direct implementation in functional devices. This can be made possible by tailoring the grains size for example. Additionally, it is known that this family of materials usually unveils several electric and magnetic ordering parameters, opening the way for the creation of potential ''cooperative'' caloric effects. Finally, to approach their theoretical limit in terms of MCE, it is necessary to develop a high fundamental understanding of the coupling between the crystalline, magnetic and electronic structures in this sort of functional materials.

**References**

[1] M. Balli, S. Jandl, P. Fournier, A. Kedous-Lebouc, Appl. Phys. Rev. 4, 021305 (2017).

[2] C. Zimm, A. Jastrab, A. Sternberg, V.K. Pecharsky, K. Gschneidner Jr., M. Osborne, I. Anderson, Adv. Cryog. Eng. 43, 1759 (1998).

[3] K. Gschneidner, V.K. Pecharsky, Int J Refrig, 31, 945 (2008).

[4] M. Balli, O. Sari, L. Zamni, C. Mahmed, and J. Forchelet, Materials Science and Engineering: B 177, 629 (2012).

[5] M. H. Phan, S. C. Yu., J. Magn. Magn. Mat. 308, 325 (2007).

[6] M. Balli, B. Roberge, P. Fournier and S. Jandl, Crystals 7, 44 (2017).

[7] M. Balli, S. Jandl, P. Fournier, M. M. Gospodinov, Appl. Phys. Lett. 104, 232402 (2014).

[8] M. Balli, S. Jandl, P. Fournier, M. M. Gospodinov, Appl. Phys. Lett. 108, 102401 (2016).

[9] N. S. Rogado, J. Li, A. W. Sleight, and M. A. Subramanian, Adv. Mater. 17, 2225 (2005).

[10] Youshun Jia, Qiang Wang, Yang Qi, Lingwei Li, J. Alloys. Comp. 726, 1132e1137 (2017)

[11] Youshun Jia, Qiang Wang, Ping Wang, Lingwei Li, Ceramics International 43, 15856 (2017).

[12] Tirthankar Chakraborty, Hariharan Nhalil, Ruchika Yadav, Aditya A. Wagh, Suja Elizabeth, J. Magn. Magn. Mat. 428, 59 (2017).

[13] J. Krishna Murthy, K Devi Chandrasekhar, Sudipta Mahana, D Topwal and A Venimadhav, J. Phys. D: Appl. Phys. 48, 355001 (2015).

[14] D. Matte, M. de Lafontaine, A. Ouellet, M. Balli, and P. Fournier, Phys. Rev. Applied 9, 054042 (2018).

[15] M. Balli, P Fournier, S Jandl, M. M. Gospodinov, J. Appl. Phys. 115, 173904 (2014).

[16] M. Balli, P. Fournier, S. Jandl, K. D. Truong, and M. M. Gospodinov, J. Appl. Phys. 116, 073907 (2014).

[17] K. I. Kobayashi, T. Kimura, H. Sawada, K. Terakura, and Y. Tokura, Nature 395, 677 (1998).

[18] K. D. Truong, M. P. Singh, S. Jandl, and P. Fournier, Phys. Rev. B 80, 134424 (2009).

[19] R. I. Dass, J.-Q. Yan, and J. B. Goodenough, Phys. Rev. B. **68**, 064415 (2003).

[20] M. P. Singh, K. D. Truong, S. Jandl, and P. Fournier, Phys. Rev. B 79, 224421 (2009).

[21] J. B. Goodenough, A. Wold, R. J. Arnott, and N. Menyuk, Phys. Rev. 124, 373 (1961).

[22] M. Sonobe and K. Asai, J. Phys. Soc. Jpn. 61, 4193 (1992).

[23] G. Blasse, J. Phys. Chem. Solids 26, 1969 (1965).

[24] C. L. Bull and P. F. McMillan, J. Sol. State Chem. 177, 2323 (2004).

[25] R. Booth, R. Fillman, H. Whitaker, A. Nag, R. Tiwari, K. Ramanujachary, J. Gopalakrishnan, and S. Lofland, Mat. Res. Bulletin 44, 1559 (2009).

[26] S. Ravi, C. Senthilkumar, Materials Letters 164, 124 (2016).

[27] K. D. Truong, J. Laverdière, M. P. Singh, S. Jandl, and P. Fournier, Phys. Rev. B. 76, 132413 (2007).

[28] J. B. Goodenough, Physical Review 100(2), 564 (1955).

[29] J. Kanamori, J. of Physics and Chem. of Solids 10(2), 87 (1959).

[30] P. W. Anderson, Theory of magnetic exchange interactions: exchange in insulators and semiconductors. Solid state physics 14, 99 (1963).

[31] J. Sanchez-Benitez, M. J. Martinez-Lope, J. A. Alonso and J LGarcia -Munoz, J. Phys.: Condens. Matter. 23, 226001 (2011).

[32] M. Retuerto, Á. Muñoz, M.-J. Martínez-Lope, J.-A Alonso, F.- J. Mompeán, M.-T. Fernández-Díaz and J. Sanchez-Benítez, Inorg. Chem. 54, 10890 (2015).

[33] T. Chakraborty, H. S Nair, H. Nhalil, K Ramesh Kumar, A. M Strydom and E. Suja, J. Phys.: Condens. Matter. 29, 025804 (2017).

[34] E. du Trémolet de Lachaisserie, Ed., *Magnétisme, fondements*, EDP, France (1999).

[35] M. Balli, D. Fruchart, D. Gignoux, and R. Zach, Appl. Phys. Lett. 95, 072509 (2009).

[36] A. Arrott and J. Noakes, Phys. Rev. Lett. 19, 786 (1967).

[37] C. Gauvin-Ndiaye, T. E. Baker, P. Karan, É. Massé, M. Balli, N. Brahiti, M. A. Eskandari, P. Fournier, A.-M. S. Tremblay, and R. Nourafkan, Phys. Rev. B. 98, 125132 (2018).

[38] A.M. Tishin and Yu. I. Spichkin, *The Magnetocaloric Effect and Its Applications*, IOP Publ., Bristol, U.K. (2003).

[39] Y. Chiba, A. Smaili, C. Mahmed, M. Balli, O. Sari, Int. J. Refrig. 37, 36 (2015).

[40] M. Balli, D. Fruchart, D. Gignoux, E. K. Hlil, S. Miraglia, P. Wolfers, J. Alloys. Comp. 442, 129 (2007).

[41] M. Balli, D. Fruchart, D. Gignoux, S. Miraglia, E. K. Hlil, and P. Wolfers, J. Magn. Magn. Mater. 316, e558-e561 (2007).

[42] J.-H. Park, E. Vescovo, H.-J. Kim, C. Kwon, R. Ramesh, T. Venkatesan, Phys. Rev. Lett. 81, 1953 (1998).

[43] L. E. Hueso, P. Sande, D.R. Miguéns, J. Rivas, F. Rivadulla, M.A. López-Quintela, J. Appl. Phys. 91, 9943 (2002).

[44] K. El Maalam, M. Balli, S. Habouti, M. Dietze, M. Hamedoun, E-K. Hlil, M. Es-Souni, A. El Kenz, A. Benyoussef, O. Mounkachi, J. Magn. Magn. Mat. 449, 25 (2018).

**Acknowledgments:**

The authors thank M. Castonguay, S. Pelletier and B. Rivard for technical support. We acknowledge the financial support from NSERC (Canada), FQRNT (Québec), CFI, CIFAR, Canada First Research Excellence Fund (Apogée Canada), Université de Sherbrooke, IRESEN (Morocco) and the International University of Rabat.

M. Balli would like to thank the Grenoble Institute of Technology (France) and especially the director of G2Elab, for having hosted him as Invited Scientist.

**Figure captions**

**Figure 1**: (a) ZFC and FC thermomagnetic curves under a field of 0.1 T for $Dy_2NiMnO_6$. (b) its 0.1 T-ZFC reverse magnetic susceptibility as a function of temperature.

**Figure 2:** Hysteresis loops experienced by $Dy_2NiMnO_6$ at two representative temperatures, 2 and 100 K.

**Figure 3**: (a) Isothermal entropy change of $Dy_2NiMnO_6$ as a function of temperature under different magnetic field variations. (b) Associated adiabatic temperature change under a magnetic field variation of 5 T.

**Figure 4**: Zero-field specific heat as a function of temperature of both $Dy_2NiMnO_6$ and $La_{1.5}Dy_{0.5}NiMnO_6$ compounds.

**Figure 5**: (a) ZFC and FC thermomagnetic curves under a field of 0.1 T for $La_{1.5}Dy_{0.5}NiMnO_6$. (b) its 0.1 T- ZFC reverse magnetic susceptibility as a function of temperature.

**Figure 6**: (a) Isothermal entropy change of $La_{1.5}Dy_{0.5}NiMnO_6$ as a function of temperature under a magnetic field change of 7 T. (b) Associated adiabatic temperature change under a magnetic field variation of 7 T.

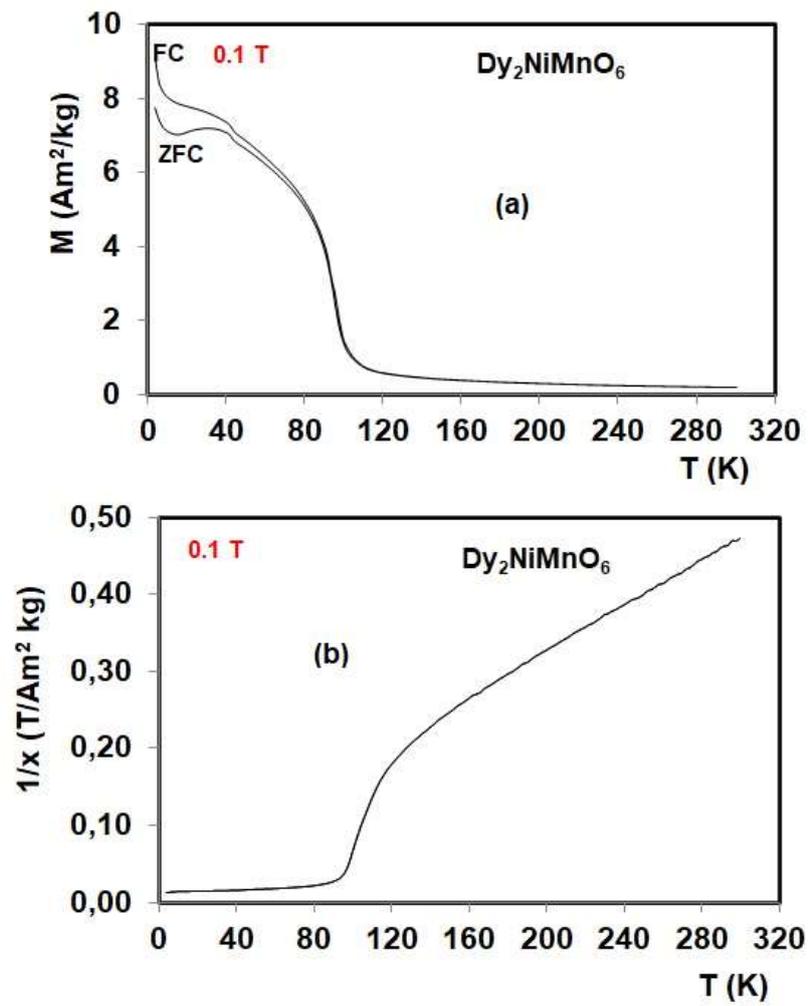

Figure 1

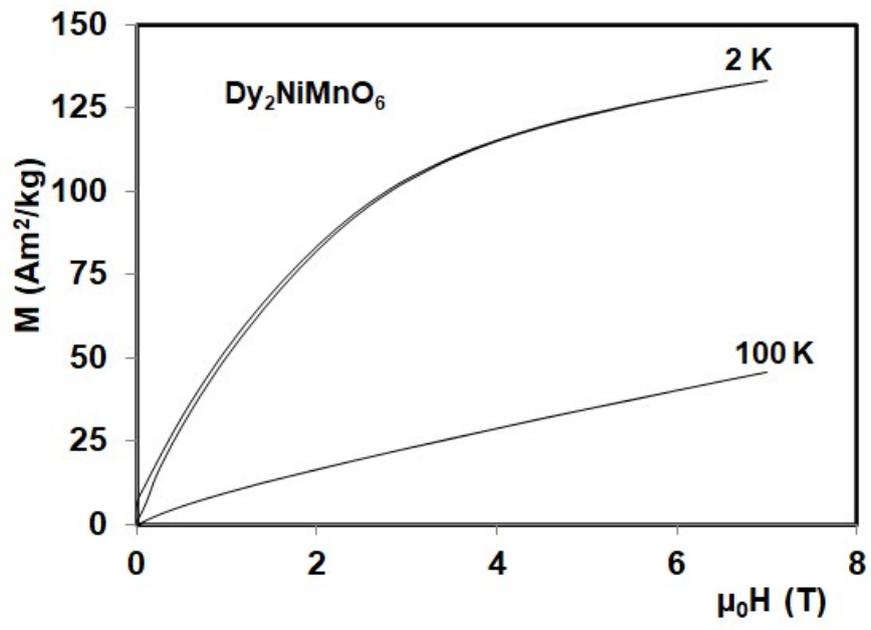

Figure 2

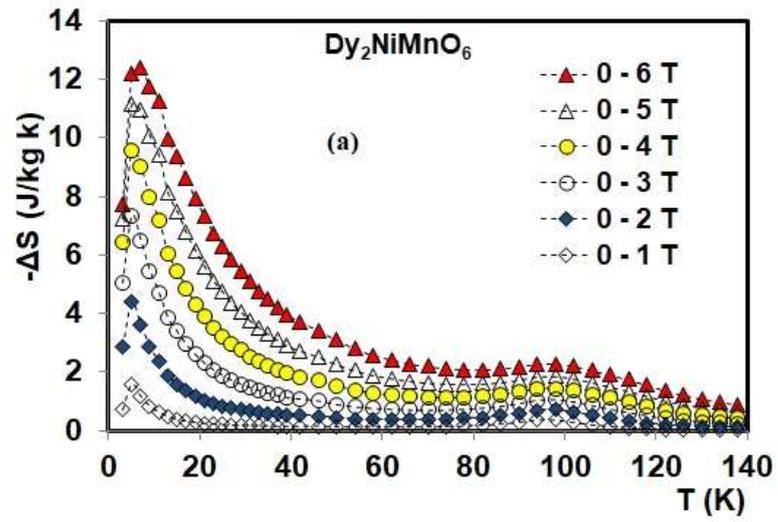

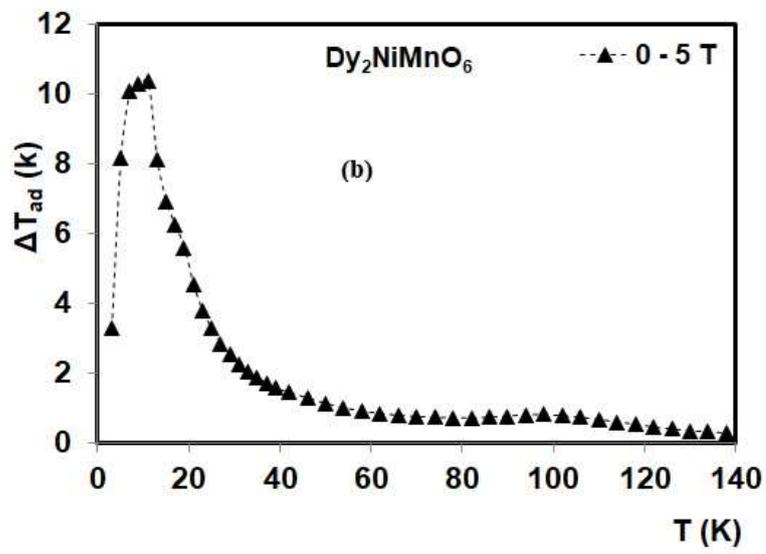

Figure 3

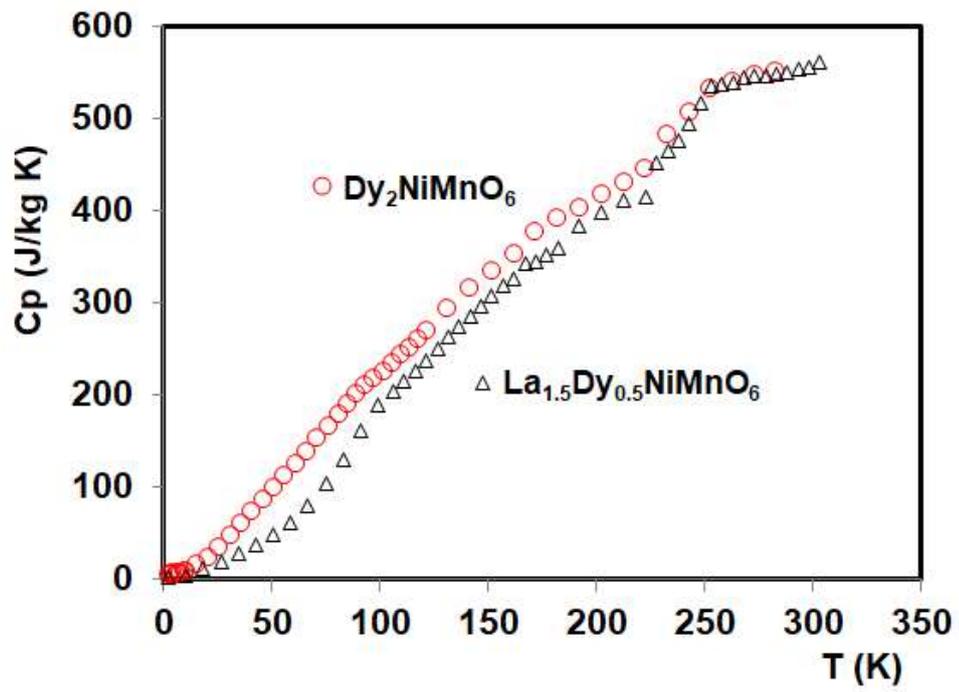

Figure 4

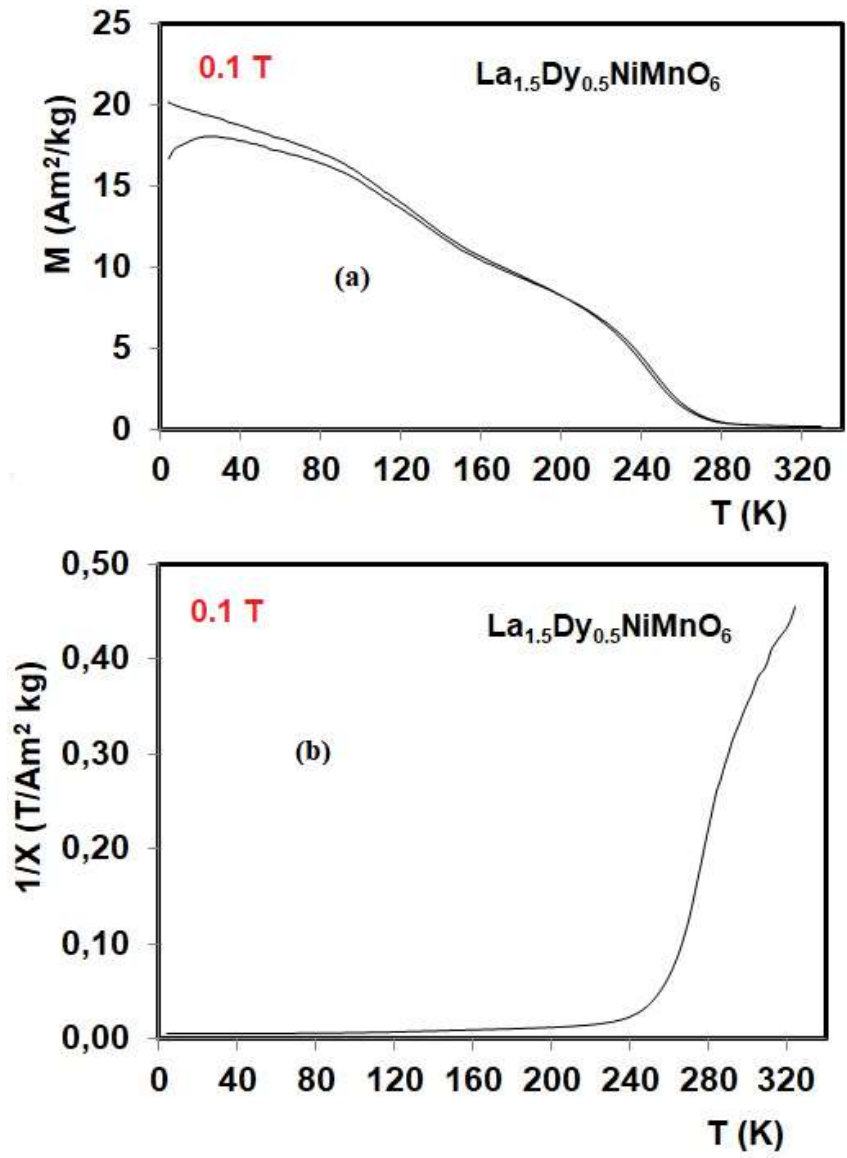

Figure 5

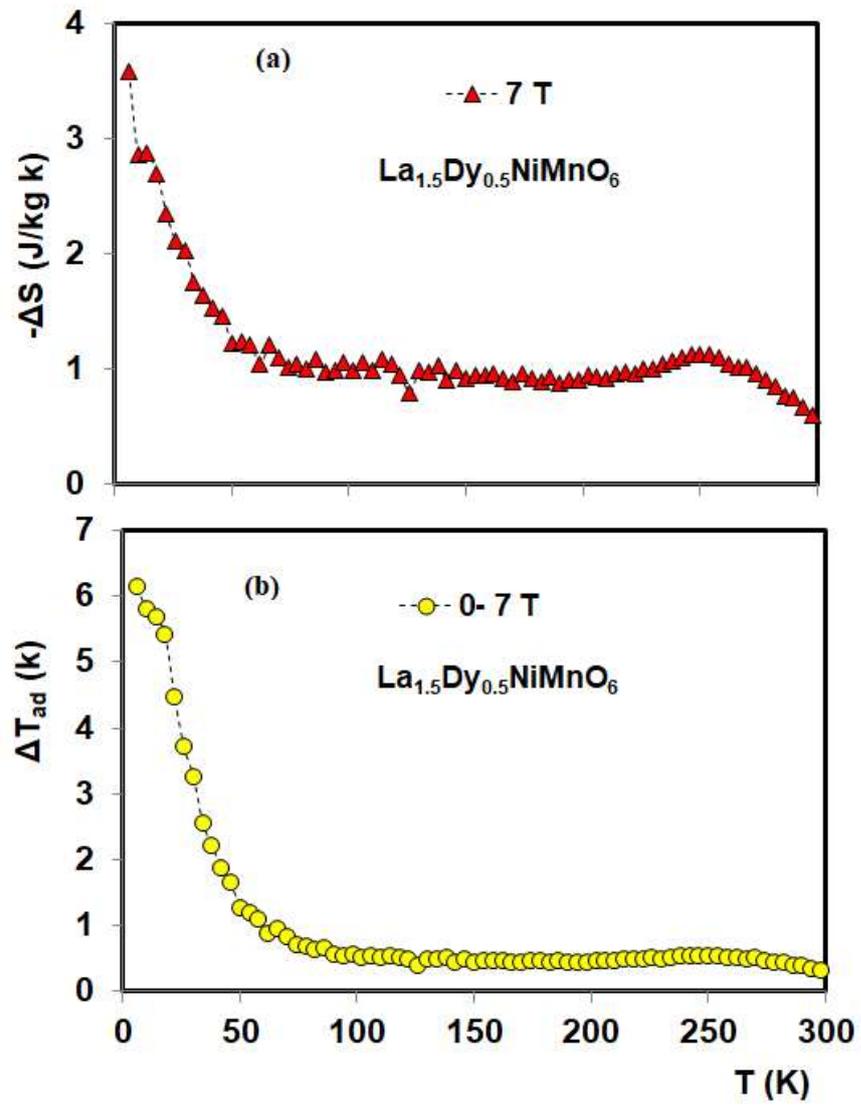

Figure 6